# Nonlogarithmic magnetization relaxation in HTSC

V.Yu.Monarkha, A.A.Shablo, and V.P.Timofeev

*B.Verkin Institute for Low Temperature Physics & Engineering National Academy of Sciences of Ukraine, 47, Lenin Ave., 61103 Kharkov, Ukraine*
*E-mail: timofeev@ilt.kharkov.ua, monarkha@mail.ru.*

**Abstract.** An experimental study of low density captured magnetic flux dynamics in YBCO single crystal samples near the superconducting phase transition point has been performed. Vortex dynamics similar to thermally assisted flux flow regime (TAFF) has been observed for the first time in such weak fields. It will be shown that Zeldov's thermoactivated creep model successfully describes the behavior of the magnetization in the YBCO single crystals in our experiments. Within the framework of this model the estimation of normalized relaxation rate was made.

In spite of a great number of theoretical and experimental papers on studying the structure and the magnetic flux dynamics in high temperature superconductors (HTSC) for the time being the full understanding of the Abricosov vortexes pinning mechanisms hasn't been achieved [1,2]. The major part of studies is focused on reveling the basic vortex grid – material interaction processes, which define the critical characteristics of the superconductor in strong magnetic fields. The low field (close to earth's magnetic field) range and the close to critical temperature range are the most unstudied for the moment.

The magnetic flux dynamics associated with flux creep and vortex hopping depends on the pinning on the sample's structural defects, and is defined by the thermoactivation energy of those metastable processes. The vortex hopping probability grows exponentially with the temperature increasing and the pinning force decreasing, therefore the HTSC microstructure and the corresponding activation energies play a major role in the magnetic flux dynamics.

In this paper we present an experimental study of low density captured magnetic flux dynamics in YBCO single crystal samples near the superconducting phase transition point. The relaxation of isothermal magnetic momentum created by captured magnetic flux (single vortexes or vortex bundles) has been registered. Contactless measurement method used provides the necessary susceptibility ($\sim 8 \cdot 10^{-11}$ A·m$^2$), acceptable thermal stabilization ($\sim 10$ mK) in the measurement chamber, and allows to completely remove all sample preparation procedures and to preserve the sample's structure intact.

As the main research object we have selected addition less orientated YBCO single crystals. Samples under study had dimensions close to 1x1 mm$^2$, and thickness from 0,015 to 0,02mm. Annealing in oxygen atmosphere necessary for obtaining maximum $T_c$ leads to a tetragonal-ortorombical structural transformation, and as a consequence, to appearance of twinning planes. It is commonly known that a structural transformation like that also takes place in polycrystals and in thin film HTSCs. To analyze the role of these defects in pinning process we have chosen single crystal YBCO samples with twinning boundaries (TB) going all along the sample's volume and parallel to the c axis and with minimal mosaicity.

HTSC are known to be type-II superconductors. The magnetic field can penetrate in the bulk of the superconductor in a form of Abrikosov's vortexes, that can "pin" on various structural defects (vacations, dislocations, dashes, intermosaic boundaries, twinning boundaries etc.) – pinning centers. Under the action of the Lorentz force that can appear due to the vortices density gradient, transport current, or thermal activation, the vortexes start to hop from one pinning center to the nearest other one, the energy dissipation appears and the superconductor transits to a resistive state. These processes connected with vortex motion define the critical current-carrying capabilities of the material under study.

In order to study magnetic flux dynamics associated with bulk pinning sites in YBCO single crystals the measurements were performed in field cooling mode: the sample was cooled in homogenous DC magnetic field. With the field disabled the vortexes are being captured by various pinning sites all over the sample. The thermally activated creep of single vortexes or vortex bundles leads to the redistribution and reduction of supercurrents, the integral magnetic momentum of the sample starts to decay and the average magnetization M starts to relax in time.

The magnetic flux dynamic research data is used to obtain important parameters of the vortex pinning mechanisms in HTSC. In the simplest case the effective pinning potential depth $U$ can be estimated from the normalized isothermal magnetization relaxation rate:

$$S(t) = \frac{1}{M_0}\left(\frac{dM}{d\ln t}\right) = -\frac{kT}{U}. \quad (1)$$

Here, $M_0$ is the initial magnetization value, which is taken to be the magnetization value in the Bin's critical state. However practically all the researches on the magnetization relaxation in HTSC were performed in high magnetic fields (~1000 Oe and more), where the major role is played by the processes in a fully formed vortex grid. In addition the experimental data obtained are very sensitive to the field orientation relative to the crystallographic planes of the sample under test, it's linear and planar defects. As it was shown in work [5], from the point of view of the collective pinning theory in weak fields non interacting vortices creep is realized. In that case the magnetic flux movement velocity, the correlation length $L_0$ and the pinning potential $U_0$ are independent from the field strength, and the measurement results are almost not sensitive to field deviation from the c axis of the YBCO crystal. Moreover the critical current value, defined by the equation between the Lorentz force and the specific pinning force is also not sensitive to field deviation.

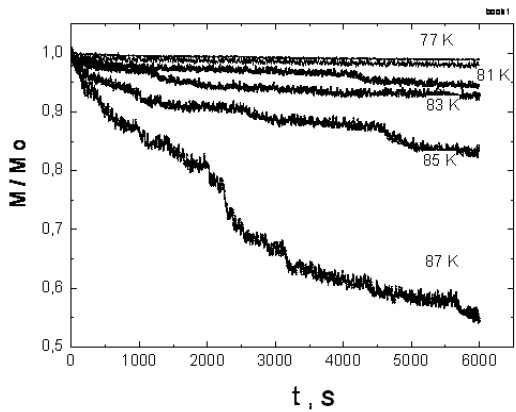

Fig.1. Normalized magnetization relaxation of YBCO single crystal in the region near Tc.

In our research we used a standard method of measuring the magnetization by registering the response of the radio-frequency SQUID-gradientometer on the magnetic moment of the sample established in homogenous magnetic field of the solenoid. While studying the particularities of superconducting phase transitions and isothermal magnetization relaxations the sample was cooled in specified small magnetic field (FC - Field Cooling method). The field then was turned off and the magnetization value change registration began. The temperature dependence of the magnetization created by the captured flux was recorded by slowly heating the sample (0.2 K/min). Then the magnetization relaxation measurements took place: the sample was cooled to the specified temperature, the field was turned off and the magnetization change in time was recorded.

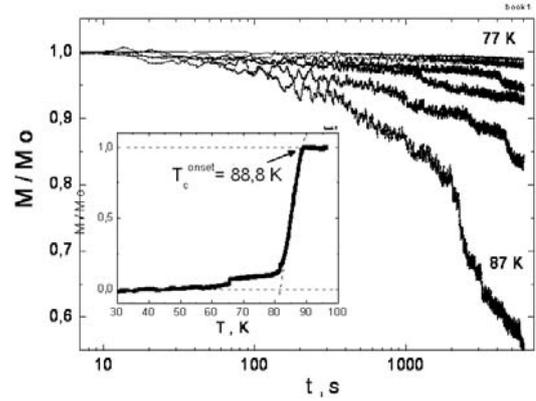

Fig.2. Normalized magnetization relaxation curves of YBCO single crystal in the region near Tc. Inset: The phase transition curve of the sample.

Figure 1 shows the dynamics of the isothermal relaxation $M(t)$, normalized to it's initial value $M_0$, for a sample with unidirectional twinning planes at different temperatures near the superconducting phase transition point. The magnetic field of the solenoid was directed along the $c$ axis of the crystal, and its value was equal to 80 A/m (~1 Oe). On the Figure 2 the same dependences are shown in a samilogarithmical scale. It's easy to notice that the linear Anderson-Kim model is incapable to describe the $M(t)$ behavior in a wide range of times and at all the temperatures.

For the temperature range $T/T_c \leq 0.8$ the sample's isothermal relaxation of magnetization has an initial area with weakly noticeable dynamics, followed by a quasilogarithmical behavior described by the Anderson-Kim model [4]. However at higher temperatures and bigger times the $M(t)$ behavior changes noticeably. Near the phase transition area and in the presence of strong thermal fluctuations the vortex dynamics is starting to become similar to thermally assisted flux flow regime (TAFF) [1]. Similar $M(t)$ behavior in such weak fields has never been observed before.

As it can be seen from the figures at small observation times the magnetization does not change, or changes very slowly. At relatively low temperatures it can be explained by exponential reduction of the thermal magnetic flux creep and by the presence of random Josephson links unsuppressed by the magnetic field in the area of the twinning boundaries [6]. The creep probability grows with temperature rising, the $M(t)$ relaxation becomes more noticeable in the time range close to a few hundred seconds after the beginning of registration. When $T$ is close to $T_c$ strong thermal fluctuations are causing giant vortex creep, and the TAFF-like dynamics is observed.

CuOx layers located in the twinning boundaries contain oxygen vacancies that cause strong influence on the suppression of the superconducting order parameter, leading to reduction of the creep activation energy for the captured vortexes. Therefore the vortex line density on the twining boundaries is higher than at the rest of the crystal. Considering that at field value ~1 Oe the intervortex distance ~$10^3$ nm is comparable with the TB period and the field penetration depth (~$10^4 - 10^5$ nm) for the current temperature range, it is reasonable to expect that all the vortices are localized on the twinning boundaries. The quasilogarithmic reduction of the magnetization is then followed by a strong thermal fluctuation area at bigger times.

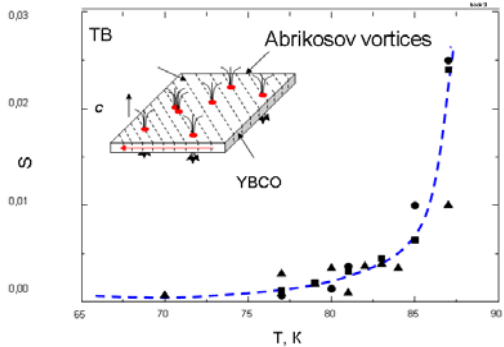

Fig.3. Normalized magnetization relaxation rate of YBCO single crystal in the region near Tc.

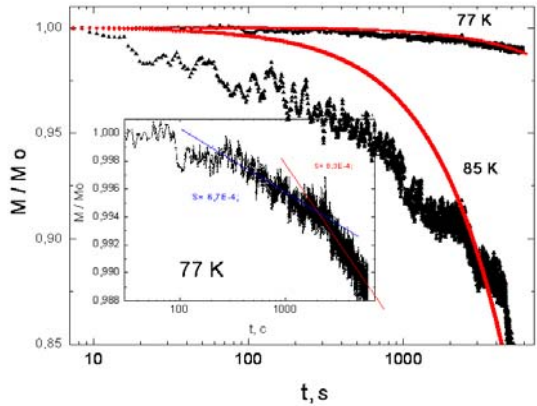

Fig.4. Relaxation of normalized magnetization for two different temperatures and the curves that describe this data according to the Zeldov's model (red). T=77 K, kT/U0=1,1E-3, t0=80s; T=85 K, kT/U0=8E-3, t0=300s. Inset: An attempt to use Anderson-Kim model.

Using the $M(t)$ relaxation results the averaged normalized relaxation speed $S$ for a wide temperature interval (70 K≤T≤87 K) has been evaluated. The value obtained does not contradict with the experimental results of the other researchers, obtained for similar samples in strong magnetic fields [3,4]. However due to a significant deviation from the logarithmical dependence the linear Anderson-Kim model (the normalized creep activation energy is $U/U_0=(1–J/J_c)$, where $U_0$ is the averaged pinning center potential in the absence of currents and thermal fluctuations) becomes unsuitable for the higher temperature range. In order to describe the experimental data we attempted to use the Zeldov's reverse logarithmical model ($U/U_0=ln(J_c/J)$ ) [7,4]. In this model the current in the superconductor depends on $U_0$ and the macroscopic time parameter $t_0$ as shown in the equations below:

$$\frac{J}{J_c} = \exp\left(-\frac{kT}{U_0}\ln\left(\frac{t}{t_0}\right)\right), \qquad (2)$$

$$S = -\frac{kT}{U_0}$$

Figure 4 shows experimental points for two different temperatures (77K and 85K) and the curves that describe this data according to the Zeldov's model. As can be seen from Fig. 4 the $M(t)$ behavior corresponds well to the model's prediction for the lower temperature band. For $T \to T_c$ the chosen model describes the experimental curves with acceptable accuracy at big observation times.

Thereby, the research on the magnetic flux dynamics captured in low fields (~1 Oe) in YBCO single crystals with unidirectional TB structure at temperatures close to $T_c$ was performed for the first time. It has been shown that the crystal structure (the TB mainly) has significant influence on the magnetization relaxation speed $S$. The possibility of nonlogatithmical $M(t)$ relaxation has been observed, and the evaluation of the effective pinning potential U has been performed using different thermoactivated creep models. We have shown that Zeldov's thermoactivated creep model successfully describes the behavior of the magnetization in the YBCO single crystals in a wide time range.

The obtained results are useful for understanding the pinning mechanisms in HTSC, and can be used on practice while constructing high sensitive superconducting devices, to minimize self-noises of the HTSC sensors and to increase sensitivity of the liquid nitrogen cooled apparatus.